\documentclass[10pt]{article}
\usepackage{mathtools,amssymb,amsthm}
\usepackage{graphicx}
\usepackage{hyperref}
\usepackage{verbatim}
\usepackage{color}
\usepackage{float}
\usepackage{subcaption}
\usepackage[section]{placeins}

\edef\restoreparindent{\parindent=\the\parindent\relax}
\usepackage{parskip}
\restoreparindent
\usepackage[normalem]{ulem}
\usepackage{fullpage}

\graphicspath{{figures/}}

\numberwithin{equation}{section}

\newcommand{\be}{\begin{equation}}
\newcommand{\ee}{\end{equation}}
\newcommand{\eq}[1]{\begin{align}#1\end{align}}

\newcommand{\bfig}{\begin{figure}}
\newcommand{\efig}{\end{figure}}

\newcommand{\cL}{\mathcal{L}}
\newcommand{\mr}{\mathbb{R}}
\newcommand{\ms}{\mathbb{S}}

\newcommand{\spac}{\Sigma}

\newcommand{\npl}{N_{pl}}

\newcommand{\cO}{\mathcal O}
\newcommand{\hW}{\widehat W}
\newcommand{\hD}{\widehat \Delta}
\newcommand{\ssee}{\mathcal S}
\newcommand{\mass}{\mathbf{m}}
\newcommand{\mx}{\mathrm{max}}

\newcommand{\hn}{\hat{n}}

\newcommand{\hphi}{\widehat \Phi}
\newcommand{\bv}{\mathbf \Phi}
\newcommand{\bk}{\mathbf k}

\newcommand{\bpp}{\mathbf p}
\newcommand{\bap}{{\bar {\mathbf p}}}

\newcommand{\bx}{\mathbf x}
\newcommand{\bu}{\mathbf \Psi}
\newcommand{\br}{\mathbf r}

\newcommand{\Righf}{I}
\newcommand{\fut}{III}
\newcommand{\Leff}{II}
\newcommand{\cM}{\mathcal{M}}

\newcommand{\cH}{\mathcal{H}}
\newcommand{\cS}{\mathcal{S}}
\newcommand{\cC}{\mathcal{C}}
\newcommand{\cB}{\mathcal{B}}
\newcommand{\rs}{{r_*}}

\newcommand{\tip}{{\tilde{p}}}

\newcommand{\tl}{\tilde{l}}
\newcommand{\nd}{{\nu_d}}
\begin{document}
\title{Spacetime Entanglement Entropy of \\ de Sitter and Black Hole Horizons}
\author{Abhishek Mathur, Sumati Surya and Nomaan X\\{\small\it Raman Research Institute, Sadashivanagar, Bangalore 560080, India}}
\date{}
\maketitle
\begin{abstract}
We calculate Sorkin's manifestly covariant entanglement entropy $\cS$ for a massive and massless minimally coupled free Gaussian scalar field for the de Sitter 
horizon and  Schwarzschild de Sitter horizons,  respectively,  in $d > 2$. In de Sitter spacetime we restrict the Bunch-Davies vacuum
in the conformal patch  to the static patch to obtain a mixed state. The finiteness of the spatial ${\mathcal
  L}^2$ norm in the static patch implies that $\cS$ is well defined for each mode.  We find that $\cS$ for this mixed
state  is  independent of the effective mass of the scalar field, and matches that of \cite{Higuchi:2018tuk}, where, a 
spatial density matrix was used to calculate the horizon entanglement entropy. Using a cut-off in the angular modes we
show that $\cS \propto A_{c}$, where $A_c$ is the area of the de Sitter cosmological horizon. Our analysis can be
carried over to the black hole and cosmological horizon in Schwarzschild de Sitter spacetime, which also has finite
spatial ${\mathcal
  L}^2$  norm in the static regions. Although the explicit form
of the modes is not known in this case, we use the boundary conditions of \cite{Qiu:2019qgp}  for a massless minimally
coupled scalar
field, to find the mode-wise $\cS_{b,c}$, where $b,c$
denote the black hole and de Sitter cosmological  horizons, respectively. As in the 
de Sitter  calculation we see that $\cS_{b,c} \propto A_{b,c}$ after taking a  cut-off in the angular modes.     
\end{abstract}

\section{Introduction}

Entanglement entropy (EE) has emerged as an important quantity in the study of quantum fields in curved spacetime. Of
particular interest is the EE of quantum fields across black hole horizons which might partially or fully account for
the Bekenstein-Hawking entropy \cite{bkls}. The importance of studying all types of horizons was pointed out by Jacobson and
Parentani~\cite{Jacobson:2003wv} who  showed that thermality and the area law are features of all causal
horizons. Cosmological horizons in de Sitter (dS) spacetime are known to have thermodynamic
properties similar to their black hole counterparts even though these
horizons are observer dependent~\cite{gibbons}. Because of the relative simplicity of these spacetimes, they provide a
useful arena to test new proposals for calculating the EE.

In most definitions of EE  one considers the entanglement of the state at a moment
of time between two  
{\it spatial} regions,  
which is restrictive in the context of quantum gravity, or even quantum fields in curved
spacetimes which may lack a preferred time. A  more global, covariant notion
of EE could be more  useful when  working with a covariant path-integral or histories based approach to  quantum
gravity.  In particular, the notion of a state at a moment of time might not survive, 
especially in theories where the manifold structure of spacetime breaks down in the deep
UV regime, as in causal set theory \cite{Bombelli:1987aa}. 

To this end, Sorkin proposed a covariant formulation of  EE for a Gaussian (free) scalar field by expressing it in
terms of the spacetime correlators or Wightman function $W(x,y)\equiv \langle\hphi(x)\hphi(y)\rangle$~\cite{ssee}.
Starting with the  pure state  $W(x,y)$ in $(\cM,g)$ whose restriction to $\cO \subset \cM$ is $W(x,y)|_{\cO}$, 
the  EE associated to this state is given by Sorkin's spacetime entanglement entropy (SSEE) formula
\be
  \ssee = \sum_\mu \mu \log |\mu|, \quad \hW|_\cO \circ \chi = i \mu  \hD \circ \chi, \quad \chi \not\in \mathrm{Ker}(i
  \hD),
 \label{ssee.eq}
\ee
where $i\hD$ is the integral operator defined via the Pauli-Jordan function $i\Delta(x,y)=[\phi(x),\phi(y)]$ and where  
\be
(\widehat{A}\circ f)(x)\equiv \int_{\cO}dV_y\,A(x,y)\,f(y).
\label{pjdef}
\ee
This expression was obtained in \cite{ssee} by noticing that for discrete spaces the field operators, which are the generators of the
  free field algebra,  can be expanded in a ``position-momentum'' basis which renders $i\hD$ into a $2\times 2$ block diagonal form
  and simultaneously diagonalises the  symmetric part of $\hW$.  Each block is a single particle system for which the
  von Neumann entropy for a Gaussian state can be calculated using results of \cite{bkls}. The expression can be then
  rearranged in terms of the eigenvalues of $\hD^{-1} \hW$. Summing over all the blocks gives the SSEE expression
  Eqn. \eqref{ssee.eq}. Apart from discreteness, an important assumption in the construction is that there is an irreducible representation
  associated with the restriction of the algebra to the region of interest. Our explicit construction in this present work is therefore an important confirmation that the
  SSEE is indeed equivalent to the von Neumann entropy in de Sitter spacetime.
 
This formula has been applied in the continuum to the $d=2$ nested causal diamonds as
well as to the causal diamond
contained in the $d=2$ cylinder spacetime,  and shown to give the expected Calabrese-Cardy
logarithmic  behaviour \cite{saravani2014spacetime,Mathur:2021zzl}. It has also been calculated in the discrete
setting, i.e., for causal sets approximated by causal diamonds in Minkowski and  de Sitter spacetimes: in $d=2$ where it
shows the expected logarithmic behaviour and in $d=4$ where it shows the expected area behaviour 
~\cite{Sorkin:2016pbz,Belenchia:2017cex,Surya_2021}. 

In this work, we present an analytic calculation of the SSEE  for de
Sitter horizons for all $d>2$ for a massive scalar field with effective mass $\mass=\sqrt{m^2+\xi R}$.  Our
calculation uses the restriction of the  Bunch-Davies vacuum in the Poincare or conformal patch of de Sitter to the
static patch. Even though $\cO$ is non-compact in the time direction, we show that the generalised eigenvalue
equation Eqn.~(\ref{ssee.eq}) can be explicitly solved mode  by mode. We find that  the SSEE 
is independent of the effective mass, which is
in agreement with  the results of Higuchi and Yamamoto \cite{Higuchi:2018tuk}\footnote{ As pointed out in
  \cite{Higuchi:2018tuk} this result differs from the EE calculated in \cite{Maldacena:2012xp}
  (see also \cite {Kanno:2014lma,Iizuka:2014rua}),  which was  found to be mass dependent. This
difference can be 
traced to the fact that the entanglement surface in \cite{Maldacena:2012xp} is not  the  de Sitter horizon but a 
superhorizon sized  surface on a hyperbolic slicing of the conformally flat de Sitter region, taken to lie close  to the future null boundary $\mathcal I^+$, with the explicit  aim of
studying  
superhorizon entanglement generated by expansion. Since the entangling regions are different
from that in our present work, it is not surprising that the results also differ.}. 
The total SSEE can be calculated using a UV cut-off
in the {angular modes} for the Bunch-Davies vacuum and is therefore proportional to the regularised  area of the horizon.  The
other $\alpha$ vacua however need an additional momentum cut-off.

The obvious  generalisation of our calculation to static black hole
and Rindler horizons is hampered by the spatial non-compactness, except in the case of Schwarzschild de Sitter
black holes. For these spacetimes, the explicit form of the modes is not known, except in $d=2$. In \cite{Qiu:2019qgp}
certain natural boundary conditions for massless minimally coupled modes were used to analyse the thermodynamic
properties of these horizons. We employ these  same boundary conditions to find the mode-wise form for the SSEE in the
static region. Introducing the cut-off in the angular modes again gives us the requisite area dependence. 

In the special case of $d=2$, the calculation can be performed explicitly, and we find that 
the SSEE is constant for both the black  hole as well as the cosmological  horizon. Thus we do not find the logarithmic
behaviour expected from the Calabrese-Cardy formula. A key difference is that  in the earlier calculations,  $\cO$ is compact and the mixed state in $\cO$ is
{\it not}  diagonal with respect to the (Sorkin-Johnston) modes in $\cO$. 
Although this is surprising, the calculations of~\cite{Higuchi:2018tuk} suggest that this is also a feature of the standard von Neumann EE in $d=2$ de Sitter spacetime.  

We organise our paper as follows. In Sec.~\ref{gf.sec} we lay out the general framework for the calculation of the
mode-dependent SSEE  for a compact region $\cO $ with respect to  a vacuum state in $\cM \supset \cO$. We find the solutions to the generalised
eigenvalue equation Eqn.~\eqref{ssee.eq} when the modes in $\cO$ are $\cL^2$ orthogonal, and the Bogoliubov
coefficients satisfy certain conditions. We then show that Eqn.~\eqref{ssee.eq} is also well posed  for static spherically symmetric spacetimes with finite
spatial extent. Assuming that the restricted vacuum $W_\cO$ is block diagonal in the modes in $\cO$ we find the general
form of the mode-wise  SSEE. In Sec.~\ref{dS.sec} we review some basics of de Sitter and  Schwarzschild de Sitter
spacetimes. In Sec.~\ref{dSssee.sec} we apply the analysis of Sec.~\ref{gf.sec} to the static patches of $d=4$ de Sitter, starting with  the Bunch-Davies vacuum  in
the conformal patch. Using an angular cut-off we show that the SSEE is proportional to the regularised de Sitter horizon
area. In Sec.~\ref{bh.sec} we  calculate the SSEE for a massless minimally coupled scalar field in  the static patches of Schwarzschild de Sitter spacetimes for $d>2$
using the boundary conditions of  \cite{Qiu:2019qgp}.  An explicit calculation of the $d=2$ case  then follows. 
We discuss the implications of our results  in Sec.~\ref{discussion.sec}. In Appendix \ref{alphabeta.sec} we
extend the de Sitter horizon calculation to the other $\alpha$ vacua in the conformal patch. We find that while the
mode-wise SSEE is still independent of the effective mass, the total SSEE needs an additional cut-off in the radial
momentum. In Appendix~\ref{ddimds.sec} we extend the $d=4$ analysis to all dimensions $d > 2$. 

\section{The SSEE: General Features} 
\label{gf.sec}

In this section we examine the SSEE generalised eigenvalue equation Eqn~\eqref{ssee.eq} using the two sets of modes in
the  regions $\cM, \cO$, where  $\cO \subset \cM$. We show that
when the modes in the subregion $\cO$ are $\mathcal L^2$ orthogonal, and the   Bogoliubov
transformations satisfy certain conditions, it is possible to find the general form for the SSEE. While not entirely
general, this covers a fairly wide  range of cases.

Let  $\{ \bv_{\bk}\} $ be the Klein-Gordon (KG) orthonormal modes in $(\cM,g)$, i.e.,
\eq{
(\bv_{\bk},\bv_{\bk'})_{\cM} = - (\bv_{\bk}^*,\bv_{\bk'}^*)_{\cM} = \delta_{\bk\bk'}\;\;\text{and}\,\,(\bv_{\bk},\bv_{\bk'}^*)_{\cM}=0,\label{kgortho.eq}
}
and $\{\bu_{\bpp}\} $ be those in the globally hyperbolic region $\cO \subset \cM$. Here $(.,.)_{\cM}$ denotes the KG inner product in $\cM$ given by
\begin{equation}
(\phi_1,\phi_2)_{\cM} = i\int_{\spac_{\cM}} d\spac^a \; \left(\phi_1^*\partial_a\phi_2 - \phi_2\partial_a\phi_1^*\right), \label{eq:kgip}
\end{equation}
where $d\spac^a$ is the volume element on the spacelike hypersurface $\spac\in\cM$ with respect to the future pointing unit normal. The corresponding Wightman function in $(\cM,g)$ is
\eq{
  W(\bx,\bx')=\sum_\bk \bv_{\bk}(\bx)\bv_{\bk}^*(\bx').\label{w.eq}
}
Since $\{\bu_{\bpp}\}$ forms a complete KG orthonormal basis in $\cO$, the restriction of $\bv_\bk$ to $\cO$ can be expressed as a linear combination of $\bu_\bpp$ modes, i.e.,
\eq{
\bv_\bk(\bx)\Big|_\cO = \sum_\bpp \left(\alpha_{\bk\bpp}\bu_\bpp(\bx) + \beta_{\bk\bpp}\bu_\bpp^*(\bx)\right),
}
where $\alpha_{\bk\bpp} = (\bu_\bpp,\bv_\bk)_\cO$ and $\beta_{\bk\bpp} = -(\bu_\bpp^*,\bv_\bk)_\cO$. The restriction
of $W(\bx,\bx')$ to  $\cO$ can thus be re-expressed in terms of $\{ \bu_{\bpp}\}$ as 
\eq{
W(\bx,\bx')\Big|_\cO  =  \sum_{\bpp\bpp'}\Big(A_{\bpp\bpp'}\bu_{\bpp}(\bx)\bu_{\bpp'}^*(\bx')+
B_{\bpp\bpp'}\bu_{\bpp}(\bx)\bu_{\bpp'}(\bx') + C_{\bpp\bpp'}\bu_{\bpp}^*(\bx)\bu_{\bpp'}^*(\bx') + D_{\bpp\bpp'}\bu_{\bpp}^*(\bx)\bu_{\bpp'}(\bx')\Big),\label{wo.eq}
}
where
\eq{
A_{\bpp\bpp'} \equiv \sum_{\bk}\alpha_{\bk\bpp}\alpha_{\bk\bpp'}^*,\;\;B_{\bpp\bpp'} \equiv \sum_{\bk}\alpha_{\bk\bpp}\beta_{\bk\bpp'}^*,\;\; C_{\bpp\bpp'} \equiv \sum_{\bk}\beta_{\bk\bpp}\alpha_{\bk\bpp'}^*,\;\; D_{\bpp\bpp'} \equiv \sum_{\bk}\beta_{\bk\bpp}\beta_{\bk\bpp'}^*.\label{abcd.eq}
}
The Pauli-Jordan function $i\Delta(\bx,\bx')=[\hat \Phi(\bx), \hat \Phi(\bx')]$ can be expanded in the modes in $\cO$
to give 
\eq{
i\Delta(\bx,\bx')=\sum_\bpp \left(\bu_{\bpp}(\bx)\bu_{\bpp}^*(\bx') - \bu_{\bpp}^*(\bx)\bu_{\bpp}(\bx')\right).\label{ido.eq}
}
The generalised eigenvalue equation for the SSEE Eqn.~(\ref{ssee.eq}) thus reduces to
\eq{&\sum_{\bpp,\bpp'}\Big(A_{\bpp\bpp'}\left<\bu_{\bpp'},\chi_\br\right>_{\cO} +
  B_{\bpp\bpp'}\left<\bu_{\bpp'}^*,\chi_\br\right>_{\cO}\Big) \bu_{\bpp}(x) + \Big(C_{\bpp\bpp'}\left<\bu_{\bpp'},\chi_\br\right>_{\cO} +
  D_{\bpp\bpp'}\left<\bu_{\bpp'}^*,\chi_\br\right>_{\cO}\Big) \bu^*_{\bpp}(x) \nonumber \\
  &= \mu_\br \sum_{\bpp} \Bigl( \left<\bu_\bpp,\chi_\br\right>_{\cO} \bu_{\bpp}(x)  -
  \left<\bu^*_\bpp,\chi_\br\right>_{\cO} \bu^*_{\bpp}(x) \Bigr),  \label{redssee.eq}
}where $\left<.,.\right>_\cO$ denotes the $\cL^2$ inner product in $\cO$ 
\eq{
\left<\phi_1,\phi_2\right>_\cO = \int_\cO dV_\bx\,\phi_1^*(\bx)\phi_2(\bx). \label{l2.eq}
}
Note that the coefficients in Eqn~\eqref{abcd.eq} can be evaluated using the relation
\eq{
W(\bx,\bx')\Big|_\cO - W^*(\bx,\bx')\Big|_\cO = i\Delta(\bx,\bx'),
\label{pbc.eq}}
so that
\eq{
A_{\bpp\bpp'}-D_{\bpp\bpp'}^*=\delta_{\bpp\bpp'}&\Rightarrow \sum_\bk \left(\alpha_{\bk\bpp}\alpha_{\bk\bpp'}^* - \beta_{\bk\bpp}^*\beta_{\bk\bpp'}\right) = \delta_{\bpp\bpp'},\label{eq:bc1}\\
B_{\bpp\bpp'}-C_{\bpp\bpp'}^*=0&\Rightarrow \sum_\bk \left(\alpha_{\bk\bpp}\beta_{\bk\bpp'}^* - \beta_{\bk\bpp}^*\alpha_{\bk\bpp'}\right) =0.\label{bc2.eq}
}

We now look for a special class of solutions of Eqn~(\ref{redssee.eq}).

To begin with we consider the case when the  $\mathcal L^2$ inner product Eqn.~(\ref{l2.eq}) is
finite  (this is the case for example if $\cO$ is compact).  We can then use the linear independence of the $\{
\bu_\bpp\}$  to obtain  the coupled equations
\eq{
\sum_{\bpp'}\Big(A_{\bpp\bpp'}\left<\bu_{\bpp'},\chi_\br\right>_{\cO} + B_{\bpp\bpp'}\left<\bu_{\bpp'}^*,\chi_\br\right>_{\cO}\Big) &= \mu_\br \left<\bu_\bpp,\chi_\br\right>_{\cO},\nonumber\\
\sum_{\bpp'}\Big(C_{\bpp\bpp'}\left<\bu_{\bpp'},\chi_\br\right>_{\cO}+ D_{\bpp\bpp'}\left<\bu_{\bpp'}^*,\chi_\br\right>_{\cO}  \Big) &= -\mu_\br \left<\bu_\bpp^*,\chi_\br\right>_{\cO}.\label{redgev.eq}
}
Next, assume that the $\{\bu_\bpp \}$  are  $\mathcal L^2$ orthogonal. Then 
\eq{\chi_\bap (\bx) = R \bu_\bap(\bx) + S \bu_\bap^*(\bx), \label{efun.eq}}
are eigenfunctions of Eqn.~\eqref{ssee.eq} if 
\eq{R A_{\bpp\bap} + S B_{\bpp\bap}  = \mu_{\bap} R \delta_{\bpp\bap} , \nonumber \\
   R C_{\bpp\bap} +  S D_{\bpp\bap} =  -\mu_{\bap} S \delta_{\bpp\bap}.  
   \label{musc.eq}}
 This has non-trivial solutions iff
 \eq{  (A_{\bpp\bap}- \mu_{\bap}\delta_{\bpp\bap} ) (D_{\bpp\bap}  +\mu_{\bap} \delta_{\bpp\bap} ) -
   B_{\bpp\bap}C_{\bpp\bap} =0. \label{RS.eq}
 }
 For $\bpp\neq \bap$ Eqns.~\eqref{eq:bc1} and \eqref{bc2.eq} this requires in particular that\footnote{This additional condition is {\it not} satisfied for example for a  causal diamond in the $d=2$ cylinder
 spacetime \cite{Mathur:2021zzl}.}\eq{|D_{\bpp\bap}|^2=|C_{\bpp\bap}|^2,\;
 \bpp\neq \bap.}
For $\bpp= \bap$, letting $A_{\bap
   \bap}=a_\bap, B_{\bap \bap}=b_\bap, C_{\bap \bap}=c_\bap, D_{\bap \bap}=d_\bap$, we see that $a_\bap, d_\bap$ are real from
 Eqn.~(\ref{abcd.eq}), so that 
 \eq{\mu_\bap^\pm=\frac{1}{2} \Biggl(1 \pm \sqrt{(1 + 2d_\bap)^2 -
     4|c_\bap|^2)}\Biggr), \label{mu.eq}}
 which is real only if \eq{(1 + 2d_\bap)^2 \geq  
   4|c_\bap|^2. \label{reality.eq}}
 This can be shown to be true using the following identity  
 \eq{\sum_{\bk} |\alpha_{\bk\bpp} - e^{i\theta} \beta_{\bk\bpp}|^2 & \geq 0 \nonumber \\
   \Rightarrow 1+2d_{\bpp} - 2 |c_{\bpp}| \cos (\theta + \theta') &\geq 0,}
 where $c_p=|c_p|e^{i \theta'}$. Taking $\theta=-\theta'$ gives us the desired relation.   
The two eigenvalues $\mu^+_\bap, \mu_\bap^-$ moreover satisfy  the relation 
 \eq{\mu^-_\bap=1-\mu^+_\bap, \label{pairs.eq}}  and therefore come in pairs $(\mu^+_\bpp,1-\mu^+_\bpp)$, as expected \cite{ssee}. 

Thus the mode-wise SSEE is 
{\eq{
\cS_\bap= \mu^+_\bap\log(|\mu^+_\bap|)+(1-\mu^+_\bap)\log(|1-\mu^+_\bap|).  \label{sp.eq}
}}
As we will see in the specific case of de Sitter and $d=2$ Schwarzschild de Sitter spacetimes, $\mu^+_\bap, \mu_\bap^- \not
\in (0,1)$ which is again consistent with the expectations of \cite{ssee}.  

In this work we are interested in  subregions  $\cO$ which are static and spherically symmetric. While non-compact in the
time direction we require them to be  compact in the spatial direction. Thus the  $\mathcal L^2$ inner product is $\delta$-function
orthogonal and not strictly finite. As we will see, this can still result in a finite $\cS_\bpp$.  In $d=4$ for example, 
\eq{\bu_{plm}(t,r,\theta,\phi) = N_{pl} R_{pl}(r) e^{-ipt} Y_{lm}(\theta, \phi), \quad p > 0, \label{ssph.eq}}
where $t\in(-\infty,\infty), \, r>0$ and $(\theta,\phi)\in \ms^2$,  $N_{pl}$ denotes an overall normalisation constant, and
$p$ is a continuous variable. Thus one has integrals 
over $p$  as well as summations over $l$ and $m$ in  Eqn.~(\ref{redgev.eq}). These modes are clearly $\cL^2$ orthogonal
since 
\eq{
\left<\bu_{plm},\bu_{p'l'm'}\right>_\cO = 2\pi |N_{pl}|^2||R_{pl}||^2\delta(p-p')\delta_{ll'}\delta_{mm'},
}
where $||R_{pl}||$ is the $\cL^2$ norm in the radial direction and finite by assumption. This $\delta$-function
orthogonality implies that for any function $\chi_\br$ (which can be expanded in terms of the complete $\{\bu_{plm}\}$
basis), both sides  of  Eqn.~(\ref{redgev.eq}) are finite.

If $\hW\Big|_\cO$ is block diagonal in the $\{\bu_{plm}\}$ basis
\eq{ A_{plmp'l'm'}=a_{plm} \delta(p-p')\delta_{ll'}\delta_{mm'}, \quad B_{plmp'l'm'}=b_{plm}
  \delta(p-p')\delta_{ll'}\delta_{mm'}, \nonumber\\
C_{plmp'l'm'}=c_{plm} \delta(p-p')\delta_{ll'}\delta_{mm'}, \quad D_{plmp'l'm'}=d_{plm}
\delta(p-p')\delta_{ll'}\delta_{mm'}.  \label{diag.eq} }
This simplifies   Eqn.~(\ref{redssee.eq}) considerably since the delta functions can be integrated  over $p'$ and
similarly,  summed  over $l',m'$.  
Using the ansatz 
\eq{\chi_{plm}(t,r,\theta,\phi) = R \bu_{plm}(t,r,\theta,\phi) + S \bu_{plm}^*(t,r,\theta,\phi), \label{efun.eq}}
for the eigenfunctions requires that Eqn.~\eqref{RS.eq} is satisfied, as before. This yields the same form for $\mu_{plm}^\pm$ as Eqn.~\eqref{mu.eq} and hence the SSEE Eqn.~\eqref{sp.eq}.

\section{Preliminaries}
\label{dS.sec}
We briefly review de Sitter and Schwarzschild de Sitter spacetimes.

de Sitter spacetime dS$_d$ in $d$ dimensions is a hyperboloid of ``radius'' $H^{-1}$  in $d+1$ dimensional  Minkowski
spacetime $\mr^{1,d}$. If $X_i$'s are the coordinates in $\mr^{1,d}$, it is the hypersurface defined by 
\be
-X_0^2+\sum_{i=1}^d X_i^2 = \frac{1}{H^2}.
\ee
We restrict our discussion to $d=4$ in what follows, since the higher dimensional generalisation is relatively
straightforward (see Appendix \ref{ddimds.sec}).
Global dS$_4$ can be parameterized\footnote{For a detailed review of coordinate systems in dS, see \cite{stromds}.} by 4 coordinates $(\tau,\theta_1,\theta_2,\theta_3)$, where $\tau$ is the global time and $\theta_i$'s are coordinates on a 3-sphere $\ms^3$. In these coordinates the metric can be written as
\be
ds^2=-d\tau^2+\frac{1}{H^2}\cosh^2(H\tau)\,d\Omega_3^2,
\ee
where, $\tau\in\mr$, $\theta_1,\theta_2\in[0,\pi]$ and $\theta_3\in[0,2\pi]$.
The causal structure of this spacetime becomes evident if we make the  coordinate transformation $\cosh(H\tau)=1/\cos T$,
so that 
\be
ds^2=\frac{1}{H^2\cos^2T}(-dT^2+d\Omega_3^2),\quad T\in\bigg(-\frac{\pi}{2},\frac{\pi}{2}\bigg).
\label{confmetric}
\ee
\bfig[h!]
    \centering
    \includegraphics[height=5cm]{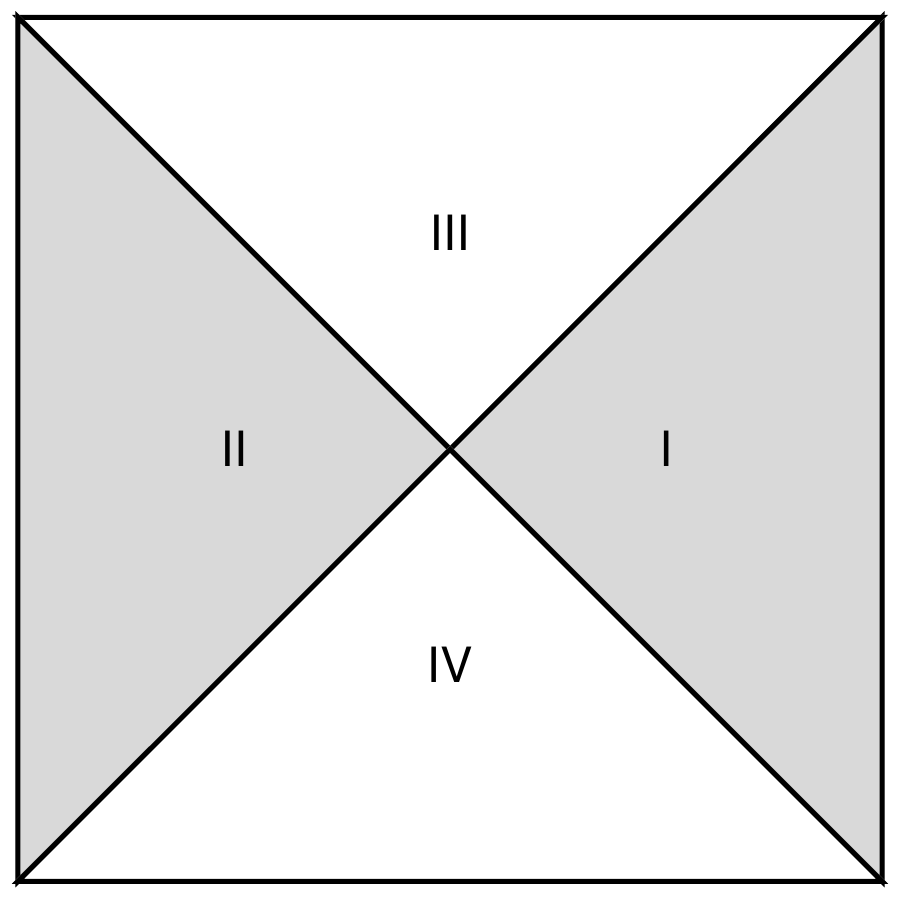}
    \caption{The Penrose diagram for dS can be deduced from the metric Eqn.~\eqref{confmetric}. Here 2 dimensions are
      suppressed so that each point represents an $\ms^2$ and each horizontal slice an $\ms^3$. dS is spatially compact,
      the left and right vertical lines correspond to $\theta_1=0,\pi$. The lower, upper horizontal
      lines correspond to $T=-\pi/2,\pi/2$ and represent the past, future null infinities respectively.} 
    \label{fig:ds}
\efig
In Fig.~\ref{fig:ds}, the region $\Righf\cup\fut$ is the right conformal patch or the Poincar\'e patch. It can be described by the metric
\be
    ds^2=\frac{1}{H^2\eta^2}\left(-d\eta^2 + dr^2 + r^2(d\theta^2+\sin^2\theta d\phi^2)\right),
\ee
where $\eta\in(-\infty,0),\;r\in[0,\infty)$ and $(\theta,\phi)\in\ms^2$. Its subregion $\Righf$ is the right static patch and is covered by the coordinates $x\in[0,1)$, $t\in\mr$, $(\theta,\phi)\in\ms^2$ which are related to the coordinates in the conformal patch by
\be
x= -\frac{r}{\eta},\quad e^{-t} = \sqrt{\eta^2-r^2},  \label{eq:conftostat}  
\ee
so that the static patch  metric is 
\be
ds^2 = \frac{1}{H^2}\left(-(1-x^2)dt^2 + \frac{dx^2}{1-x^2} + x^2d\Omega_2^2\right).
\ee

\begin{figure}[h!]
\centering
\includegraphics[height=5cm]{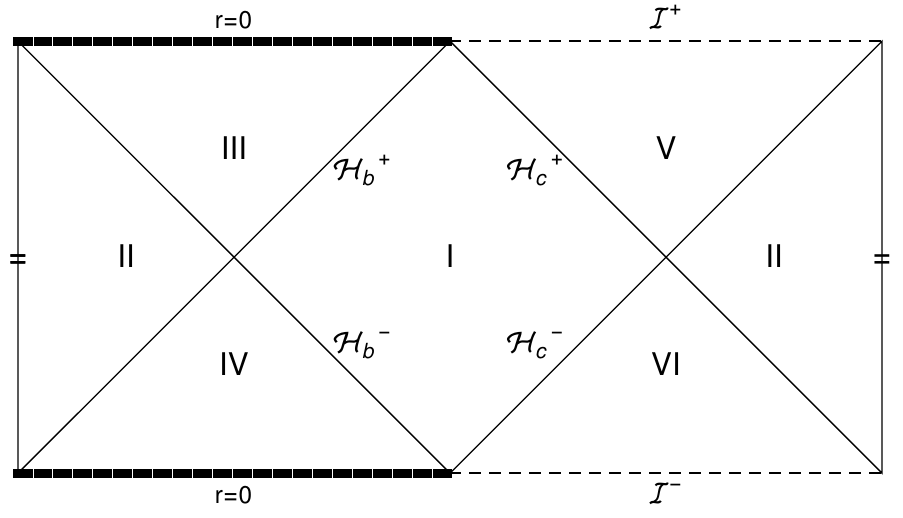}
\caption{{The Penrose diagram for the $d>2$ Schwarzschild de Sitter spacetime where each point represents an
    $\ms^{d-2}$ and each horizontal slice represents an $\ms^{d-2}\times\ms^1$. Region $I$ and $II$ are the static patches,
    and $\cH_b^{\pm}$ and $\cH_c^{\pm}$ are the black hole and the cosmological horizons respectively.}}
\label{fig:sds}
\end{figure}

We now turn to the Schwarzschild-de Sitter spacetime, whose conformal diagram is shown in Fig \ref{fig:sds}. 
It has two sets of horizons each  in regions
$I$ and $II$: the cosmological horizons $\cH_c^\pm$  and the black hole horizons $\cH_b^\pm$,  with the latter contained ``inside'' the
former.

In either of the static patches,  $I$ or $II$, the  metric of the Schwarzschild de Sitter spacetime is 
\eq{
ds^2 &= -f(r)dt^2+\frac{dr^2}{f(r)}+r^2(d\theta^2+\sin^2(\theta)d\phi^2),\quad f(r) = 1-\frac{2M}{r}-H^2r^2 \label{sdssc.eq}\\
&= -f(r)dudv +r^2(d\theta^2+\sin^2(\theta)d\phi^2),\label{sdstc.eq}
}
where $H$ is the Hubble constant and $M$ is the mass of the black hole $r\in(r_b,r_c)$, $t\in(-\infty,\infty)$ and
$(\theta,\phi)\in \ms^2$. Here $r_b<r_c$ are the real and positive solutions of $f(r)=0$, which correspond to the black
hole and the cosmological horizons $\cH_b^\pm , \cH_c^\pm$ respectively. They are related to $M$ and $H$ as
\eq{
M= \frac{r_br_c(r_b+r_c)}{2(r_b^2+r_c^2+r_br_c)},\quad H^2=\frac{1}{r_b^2+r_c^2+r_br_c}.
}
$u,v\in(-\infty,\infty)$ are the light-cone coordinates defined
as $u=t-\rs$ and $v=t+\rs$, where $d\rs=\frac{dr}{f(r)}$ \cite{Anderson:2020dim}.

As in the Schwarzschild spacetime, there is a Kruskal extension beyond the black hole and  the
cosmological horizon, given respectively by 
\eq{
U_b = -\kappa_b^{-1}e^{-\kappa_b u}\quad\text{and}\quad V_b = \kappa_b^{-1}e^{\kappa_b v},\label{schtokrusU.eq}\\ 
\quad U_c = \kappa_c^{-1}e^{\kappa_c u}\quad\text{and}\quad V_c = -\kappa_c^{-1}e^{-\kappa_c v},
\label{schtokrusV.eq}}
where $\kappa_b$ and $\kappa_c$ are the surface gravity of the black hole and the cosmological horizon
respectively \cite{Anderson:2020dim}
\eq{\kappa_b= \frac{H^2}{2r_b}(r_c-r_b)(r_c+2r_b), \quad \kappa_c=\frac{H^2}{2r_c}(r_c-r_b)(2r_c+r_b) . }
In these coordinates, the spacetime metrics in  region $\cB \equiv I\cup II\cup III\cup IV$ and  $\cC\equiv I\cup II\cup V\cup VI$ in
Fig.~\ref{fig:sds}  are, respectively 
\eq{ds_{\cB}^2 &= -f(r)e^{-2\kappa_b\rs}dU_bdV_b + r^2(d\theta^2+\sin^2(\theta)d\phi^2), \\ ds_{\cC}^2 & =
  -f(r)e^{2\kappa_c\rs}dU_cdV_c + r^2(d\theta^2+\sin^2(\theta)d\phi^2).}

\section{SSEE for Cosmological and Black Hole Horizons}\label{sseedsbh.sec}
We now calculate the SSEE for the static regions in both the de Sitter and Schwarzschild de Sitter spacetimes. In both cases, since the static region is spatially finite, Eqn.~\eqref{redssee.eq} is well-defined.

\subsection{The SSEE in the de Sitter Static Patch}\label{dSssee.sec}

We are interested in finding the entanglement across the intersection sphere $\ms^2\simeq \Righf\cap\Leff$ in
Fig.~\ref{fig:ds}. The associated sub-region $\cO$ of interest to the SSEE calculation is therefore the right or left
static region. Without loss of generality we henceforth pick the right static region $R$ and take as the larger region $\cM \supset \cO$  the conformal patch
$\Righf\cup\fut$

In the larger region $\Righf\cup\fut$, we have a well known, complete, Klein-Gordon orthonormal set of modes for a free scalar field of effective mass $\mass=\sqrt{m^2+\xi R}$ (where $\xi=1/6$ and $R$ is the Ricci scalar, which is a constant for de Sitter spacetimes) called the Bunch-Davies modes~\cite{Bunch:1978yq}. These are given by $\bv_{klm}\equiv\varphi_{kl}(\eta,r)Y_{lm}(\theta,\phi)$, where
\be
\varphi_{kl}(\eta,r) = \frac{H e^{-\frac{i\pi}{2}(l+\frac{1}{2})}}{\sqrt{2k}}(-k\eta)^{\frac{3}{2}}e^{\frac{i\nu\pi}{2}}H_\nu^{(1)}(-k\eta)j_l(kr). \label{eq:bdmodes}
\ee
Here $k\in{\mathbb{R}}^+,\,l\in\{0,1,2,\dots\}$, $m\in\{-l,..,0,..,l\}$, $Y_{lm}$s are the spherical harmonics on ${\mathbb{S}^2}$, $j_l$ is the spherical Bessel function and $H_\nu^{(1)}$ is the Hankel function of the first kind with
\be
    \nu= \sqrt{\frac{9}{4}-\frac{\mass^2}{H^2}}, 
    \label{nudef}
    \ee
and satisfies the plane-wave behaviour expected at late times.

In the region $\Righf$ we have a complete set of Klein-Gordon orthonormal modes \cite{Higuchi:1986ww} given by $\bu_{plm}\equiv\psi_{pl}(t,x)Y_{lm}(\theta,\phi)$ where
\be
    \psi_{pl}(t,x) \equiv \sqrt{2\sinh(\pi p)}\,N_{pl} \,U_{pl}(x)e^{-ipt},\quad p\in\mr^+, \label{eq:stat_modes}
\ee
where
\be
    N_{pl}=\frac{H}{2\sqrt{2}\pi\Gamma(l+\frac{3}{2})}\Gamma\left(\frac{\frac{3}{2}+l-ip+\nu}{2}\right)\Gamma\left(\frac{\frac{3}{2}+l-ip-\nu}{2}\right),
    \label{npldef}
\ee
and
\be
    U_{pl}(x)=x^l(1-x^2)^{\frac{-ip}{2}}{_2F_1}\left(\frac{\frac{3}{2}+l-ip+\nu}{2},\frac{\frac{3}{2}+l-ip-\nu}{2},l+\frac{3}{2};x^2\right).
    \label{Upldef}
\ee
As shown in \cite{Higuchi:1986ww}, 
\begin{enumerate}
	\item $U_{pl}(x)=U_{-pl}(x)=U^*_{pl}(x)$, which can be shown using an identity of the Hypergeometric function i.e., ${_2F_1}(a,b,c;z)=(1-z)^{c-a-b}{_2F_1}(c-a,c-b,c;z)$.
	\item $N_{pl}=N^*_{-pl}$, which comes from the identity $\Gamma^*(z)=\Gamma(z^*)$.
        \end{enumerate}
As discussed in Sec.~\ref{gf.sec}, being static and spherically symmetric, the  $\bu_{plm}$  modes are also $\mathcal
L^2$ orthogonal in $\Righf$.

We now proceed to obtain the SSEE for the sub-region $\Righf$ with respect to the Bunch-Davies vacuum in the right conformal patch $\Righf\cup\fut$. As suggested in Sec.~\ref{gf.sec}, we begin by demonstrating that the Bogoliubov coefficients between the Bunch-Davies modes
$\bv_{klm}$ and the static modes $\bu_{plm}$ in $\Righf$ satisfy the criteria Eqn.~\eqref{diag.eq}.

Since the $(\theta,\phi)$ dependence of both sets of modes is given by $Y_{lm}(\theta,\phi)$, which themselves are
linearly independent in $\ms^2$, the Bogoliubov transformation is non-trivial only between $\varphi_{kl}$ and
$\psi_{pl}$ for each $l,m$, i.e., 
\begin{equation}
\varphi_{kl}(\eta,r) = \int_0^\infty dp\,\left(\alpha_{kp}\psi_{pl}(t,x) + \beta_{kp}\psi_{pl}^*(t,x)\right)\label{eq:vtophi}.
\end{equation}
Instead of using the Klein-Gordon inner product to calculate $\alpha_{kp}$ and
$\beta_{kp}$, we can use the $\cL^2$ orthogonality of the $\bu_{plm}$ modes as well as the $\cL^2$ inner product of $\bv_{klm}$ and $\bu_{plm}$ in $\Righf$, so that 
\begin{equation}
\alpha_{kp} = \frac{1}{n_p}\left<\bu_{plm},\bv_{klm}\right>_\Righf\quad\text{and}\quad \beta_{kp} = \frac{1}{n_p}\left<\bu_{pl-m}^*,\bv_{klm}\right>_\Righf,\label{eq:pqip}
\end{equation}
with $n_p = 4\pi\sinh(\pi p)|N_{pl}|^2||U_{pl}||^2$ being the $\cL^2$ norm of the $\bu_{plm}$ modes.
The identity \cite{magnus}
\eq{
\int_0^\infty \!\!dz\, z^\lambda H_\nu^{(1)}(az)J_\mu(bz)
&=a^{-\lambda-1}e^{i\frac{\pi}{2}(\lambda-\nu+\mu)}\frac{2^\lambda(b/a)^\mu}{\pi\Gamma(\mu+1)}\Gamma\left(\frac{\lambda+\nu+\mu+1}{2}\right)\Gamma\left(\frac{\lambda-\nu+\mu+1}{2}\right)\nonumber\\
&\times_2F_1\left(\frac{\lambda+\nu+\mu+1}{2},\frac{\lambda-\nu+\mu+1}{2},\mu+1,\left(\frac{b}{a}\right)^2\right), 
\nonumber \\  &\quad\quad\quad\quad\quad\quad\mathrm{Re}(-i(a\pm b))>0, \, \, \mathrm{Re}(\mu+\lambda+1\pm\nu)>0,
}
 can be used as in  \cite{Higuchi:2018tuk}, to show that
\eq{
\frac{1}{\sqrt{2\pi}}\int_0^\infty dk\;k^{-ip-\frac{1}{2}}\varphi_{kl}(\eta,r) &= 2^{-ip}e^{\frac{\pi p}{2}}N_{pl}\left(\eta^2-r^2\right)^{\frac{ip}{2}}U_{pl}\left(-\frac{r}{\eta}\right)\nonumber\\
&= 2^{-ip}e^{\frac{\pi p}{2}}N_{pl}U_{pl}(x)e^{-ipt},
}
where we have substituted $\lambda=-ip$, $\mu=l+1/2$, $a=-\eta$ and $b=r$. Inverting the above, 
\eq{
\varphi_{kl}(\eta,r) = \frac{1}{\sqrt{2\pi}}\int_{-\infty}^\infty dp\,2^{-ip}k^{ip-\frac{1}{2}}e^{\frac{\pi p}{2}}N_{pl}U_{pl}(x)e^{-ipt},\label{eq:vtor}
}
using which 
\eq{
\alpha_{kp} &= \frac{1}{\sqrt{2\pi}n_p}\int_{-\infty}^\infty dp'\,2^{-ip'}k^{ip'-\frac{1}{2}}e^{\frac{\pi
    p'}{2}}\sqrt{2\sinh(\pi p)}N_{p'l}N_{pl}^*
\int_0^1 dx\,x^2 U_{p'l}(x)U_{pl}(x) \int_{-\infty}^\infty dt e^{-i(p'-p)t}\nonumber\\
&=\frac{2^{-ip}k^{ip-\frac{1}{2}}}{\sqrt{2\pi(1-e^{-2\pi p})}},\label{eq:alpkp}\\
\beta_{kp} &= \frac{1}{\sqrt{2\pi}n_p}\int_{-\infty}^\infty dp'\,2^{ip'}k^{-ip'-\frac{1}{2}}e^{\frac{-\pi p'}{2}}\sqrt{2\sinh(\pi p)}N_{p'l}^*N_{pl}
\int_0^1 dx\,x^2 U_{p'l}(x)U_{pl}(x)\int_{-\infty}^\infty dt e^{i(p'-p)t}\nonumber\\
&= \frac{2^{ip}k^{-ip-\frac{1}{2}}}{\sqrt{2\pi(e^{2\pi p}-1)}}.\label{eq:betkp}
}
Notice that the Hubble constant $H$ drops out of these coefficients. Further calculation shows that 
\begin{equation}
A_{pp'}=\frac{\delta(p-p')}{1-e^{-2\pi p}},\;\;D_{pp'}=\frac{\delta(p-p')}{e^{2\pi p}-1}\;\;\text{and}\;\;B_{pp'}=C_{pp'}=0.\label{eq:cresult}
\end{equation}
This is precisely of the form Eqn.~\ref{diag.eq}, with  $b_{p}=c_{p}=0$, where we have suppressed the $l,m$ indices.
Using the ansatz 
\eq{\chi_{p}^+(t,r) = u_{p}(t,r), \quad \chi_{p}^-(t,r)  =  u_{p}^*(t,r), \label{efun.eq}}
for the generalised eigenfunctions of Eqn.~\eqref{redssee.eq}, we  find the generalised eigenvalues 
\be
\mu^{+}_{p} = \frac{1}{1-e^{-2\pi p}}\quad\text{and}\quad \mu^{-}_{p} =- \, \frac{e^{-2\pi p}}{1-e^{-2\pi p}},
\ee
respectively, for each $p\in\mr^+$.  {Note that $\mu^{+}_{p} \in [1, \infty)$ and  $\mu^{-}_{p} \in (-\infty, 0]$,
  as expected for the SSEE \cite{ssee}.} For a given $p, l, m$ the mode-wise SSEE  is therefore 
\be
\cS_p = -\log(1-e^{-2\pi p}) - \frac{e^{-2\pi p}}{1-e^{-2\pi p}}\log e^{-2\pi p}, \label{eq:hy}
\ee
which agrees with the result of \cite{Higuchi:2018tuk}. Since there is no dependence on $l,m$ there is an infinite degeneracy coming
from  the angular modes $l\in\{0,1,\ldots\}$ and $m\in\{-l,\ldots,0,\ldots,l\}$.

In order to calculate the total SSEE, therefore,  one has to sum over the $l,m$ and integrate over $p \in (0,\infty)$.  For the  Bunch-Davies vacuum the integral  over $p$ is finite. However in the absence of a cutoff in $l$, there is an
infinite degeneracy for every $p$ coming from the angular modes which  leads to an infinite factor in
the total entropy. This ``density of states''  for a given $p$ can be regulated by
introducing a cut-off  $l_\mx$, so that
\eq{
\cS =  \sum_{l=0}^{l_\mx} \sum_{m=-l}^l \int dp \, \cS_p =
\frac{\pi}{6}(l_\mx+1)^2 \simeq \frac{\pi}{6}l_\mx^2\label{eq:sseebdvac}
}
for $l_\mx >>1 $.  $l_\mx$ can in turn be interpreted as coming from the regularised area of the de Sitter horizon $\Righf\cap\Leff
  \simeq \ms^2$.  Let us for the moment suppress one of the angular variables so that the the modes on 
 an $\ms^1$ of radius $H^{-1}$  are $e^{im\phi}$. A UV cut-off  $m_\mx$  corresponds to a minimal angular
   scale $\Delta \phi=2\pi/m_\mx$ and hence a length cut-off  $\ell_c$, where $m_\mx=\frac{2 \pi}{Hl_c}$. Thus $m_\mx$
   is the  circumference of the $\ms^1$ in units of the cut-off.  A similar argument carries over to
   $\ms^2$, where we first place $\theta$ and $\phi$ on similar footing by  writing the spherical harmonics as a Fourier series\cite{Hofsommer1960}
\eq{
Y_{lm}(\theta,\phi)\propto P_l^m(\cos\theta)e^{im\phi}=\sum_{j=-l}^l\tilde{P}_{jl}^m e^{ij\theta}e^{im\phi}. 
}
Thus, we again have the angular cut-offs $\Delta\theta=2\pi/l_\mx, \Delta \phi=2 \pi/l_\mx$, so that $\l_\mx^2 = 4
\pi^2/\Delta \theta \Delta \phi$. For large $l_\mx$ the planar limit of the  region subtended by the solid angle $\Delta
\Omega= \sin \theta \Delta\theta\Delta\phi$ on $\ms^2$ can be taken near the equator, $\theta = 
\pi/2-\epsilon$, where the metric is nearly flat in $(\theta, \phi)$ coordinates: $ds^2\simeq d\epsilon^2 +
d\phi^2$. Thus,  $l_\mx^2\propto 1/\Delta\Omega$ and therefore
\eq{
\cS\propto \frac{A_{c}}{l_c^2}
}
where we have defined a fundamental cut-off  $l_c^2=H^{-2} \Delta\Omega$ and $A_{c}=4 \pi H^{-2}$ is the area of the de
Sitter cosmological horizon. 

As shown in Appendix~\ref{alphabeta.sec} for all  the other $\alpha$ vacua, the integral $\int_{0}^p dp \,\cS_p$ is not
finite. This necessitates an additional  cut-off $p_\mx$. The extension to general 
$d>2$ is shown in Appendix~\ref{ddimds.sec}. 

\subsection{The SSEE of Schwarzschild de Sitter Horizons}\label{bh.sec}

In the Schwarzschild de Sitter spacetime, regions
$I$ and $II$ are static  and spherically  symmetric, which means that the massless scalar field modes are of the form
Eqn.~\eqref{ssph.eq}. What is important for our analysis is that the spacetime is spatially bounded so that the
calculations of Sec.~\ref{gf.sec} can be applied to this case. Without loss of generality we will work with region $I$ to calculate  its
SSEE. 

Although our focus is the $d>2$ case, we begin by suppressing  the angular dependence and considering the $d=2$ case first.
The massless, Klein-Gordon orthonormal scalar field modes are then simply the plane waves in $I$
\eq{
\bu_{p}^{(1)}(u) = \frac{1}{\sqrt{4\pi p}}e^{-ipu}\quad\text{and}\quad \bu_{p}^{(2)}(v) = \frac{1}{\sqrt{4\pi p}}e^{-ipv},\;p>0,
}  as well as in regions $\cB$ and $\cC$ 
\eq{
\bv_{k}^{(1)}(U) = \frac{1}{\sqrt{4\pi k}}e^{-ikU}\quad\text{and}\quad \bv_{k}^{(2)}(V) = \frac{1}{\sqrt{4\pi k}}e^{-ikV},\;k>0,
}
where we have suppressed the $b,c$ indices in $(U_{b,c},V_{b,c}) $ for simplicity. Note that the modes in region $I$
are static, and of the form Eqn.~(\ref{ssph.eq}), with $l\in \{0,1 \}$ representing the left and right movers. This
means that the radial part is $\cL^2$, which ensures finiteness of Eqn.~\eqref{redssee.eq}.

The restriction of  $\bv_{k}^{(1,2)}$ to region I can be written in terms of $\bu_p^{(1,2)}$ as
\eq{
\bv_{k}^{(1,2)} = \int_0^\infty dp\, \left(\alpha_{kp}^{(1,2)}\bu_{p}^{(1,2)} + \beta_{kp}^{(1,2)}\bu_{p}^{(1,2)*}\right),
}
where
\eq{
\alpha_{kp}^{(1)}=\frac{1}{2\pi}\sqrt{\frac{p}{k}}\int_{-\infty}^\infty du\, e^{ipu}e^{-ikU} &= \frac{1}{2\pi
  \kappa}\sqrt{\frac{p}{k}}\left(\frac{k}{\kappa}\right)^{i\frac{p}{\kappa}}e^{\frac{\pi p}{2 \kappa}}\Gamma\left(-i\frac{p}{\kappa}\right)\label{eq:alpha2d},\\
\beta_{kp}^{(1)}=\frac{1}{2\pi}\sqrt{\frac{p}{k}}\int_{-\infty}^\infty du\, e^{-ipu}e^{-ikU}
&=\frac{1}{2\pi\kappa}\sqrt{\frac{p}{k}}\left(\frac{k}{\kappa}\right)^{-i\frac{p}{\kappa}}e^{-\frac{\pi p}{2 \kappa}}\Gamma\left(i\frac{p}{\kappa}\right)\label{eq:beta2d},
}
$\alpha_{kp}^{(2)}=\alpha_{kp}^{(1)*}$ and $\beta_{kp}^{(2)}=\beta_{kp}^{(1)*}$
for the black hole horizon. For the cosmological horizon, they  are complex conjugates of
Eqns.~\eqref{eq:alpha2d} and \eqref{eq:beta2d}. 
Thus, we find that for both $(1,2)$ modes,
\eq{
A_{pp'}=a_p\delta(p-p'),\;\;D_{pp'}=d_p\delta(p-p')\;\;\text{and}\;B_{pp'}=C_{pp'}=0, \label{2ddiag.eq}
}
with 
\eq{
  a_p=\frac{1}{1-e^{-2\pi \frac{p}{\kappa}}}\quad\text{and}\quad d_p=\frac{e^{-2\pi \frac{p}{\kappa}}}{1-e^{-2\pi \frac{p}{\kappa}}}
  .}
Using Eqn.~\eqref{mu.eq} and the dimension-free $\tip\equiv p\kappa^{-1}$, 
we see that 
\eq{
  \cS_\tip
  = -\log\left(1-e^{-2\pi \tip}\right)
  - \frac{e^{-2\pi \tip}}{1-e^{-2\pi \tip}}\log\left(e^{-2\pi \tip}\right).
}
The total entropy is then 
\eq{
\cS=2\int_0^\infty d\tip \cS_\tip =-\frac{2}{\pi }\int_0^1 dz \frac{\log(z)}{1-z} = \frac{\pi}{3},\label{eq:rindssee}
}
where $z=e^{-2\pi \tip}$ and the factor of two comes from the fact that the total entropy is the sum of the entropy of
the  $(1,2)$ modes.  $\cS$ is therefore the same for both horizons.

{We now consider the $d=4$ case by using the boundary conditions of \cite{Qiu:2019qgp}.  As mentioned earlier, the full
  modes are not known, but the boundary conditions suffice to calculate the Bogoliubov coefficients. For our purposes it
  suffices to use the past boundary conditions, since this defines the Klein Gordon norm on the 
limiting initial null surface $\cH_b^- \cup \cH_c^-$ in Region I.  
For the static patch modes, which are of the form Eqn.~\ref{ssph.eq}, these  boundary conditions are 
\eq{
\bu_{plm} = \begin{cases}\frac{1}{\sqrt{4\pi p}r_b}e^{-ipu}Y_{lm}(\theta,\phi)\quad &\text{on} \;\cH_b^-\\ 0 \quad &\text{on}\; \cH_c^-
\end{cases},\quad p>0,\label{statmodes4d.eq}
}
while for the Kruskal modes across the black hole horizon, they are 
\eq{
\bv_{klm} = \begin{cases}\frac{1}{\sqrt{4\pi k}r_b}e^{-ikU_b}Y_{lm}(\theta,\phi)\quad &\text{on} \;\cH_b^-\\ 0 \quad &\text{on}\; \cH_c^-
\end{cases},\quad k>0,\label{krusmodes4d.eq}
}
where $U_b$ is related to $u$ as in Eqn.~\eqref{schtokrusU.eq} \cite{Qiu:2019qgp}. Note that our normalisation differs
  from that of \cite{Qiu:2019qgp} and comes from the KG norm on $\cH_b^-\cup \cH_c^-$ or equivalently $\cH_b^-$ for these
  boundary conditions. The factor $r_b^{-1}$ is dimension
  dependent and comes from the 
  normalisation of the modes along $\cH_b^-$ where $r=r_b$, and the angular measure is $r_b^2 d\Omega$. Thus for any $d>2$,
  one must include a  factor $r_b^{-\frac{d-2}{2}}$ to normalise the modes. Importantly,  these boundary conditions 
    are not appropriate for  $d=2$, since the left and right movers are independent in that case. Setting the modes to
    zero on $\cH_c^-$ in $d=2$ would thus lead to an incomplete set of modes in region I. This is not the case for
    $d>2$, where there is a ``mixing'' or scattering of the left movers on $\cH_b^-$ in region I.  
  
Since the modes vanish along $\cH_c^-$, the KG norm can be defined using only $\cH_b^-$ in region $I$ of
Fig.~\ref{fig:sds}, where $u\in(-\infty,\infty)$ and $U_b\in(-\infty,0)$. As in the de Sitter calculation, the angular
modes for $\bv_{klm}$ and  $\bu_{plm}$  are the same, so that the calculation reduces to the $d=2$ case described
above, with the Bogoliubov coefficients given by Eqn.~\eqref{eq:alpha2d} and \eqref{eq:beta2d}. Note that unlike  $d=2$,
there is only {\it one} set of complete modes, which corresponds in our case to the set $(1)$.

Thus, the SSEE  is given by the $d=2$ SSEE for one mode,  multiplied as in the de Sitter case, by the angular cut-off
term, $(l_\mx+1)^2$ coming from the degeneracy of the generalised eigenfunctions. A similar calculation can be done for
the cosmological horizon, so that we have  
\eq{\cS_\cB\propto  \frac{A_\cB}{\ell_c^2}, \quad \cS_\cC\propto  \frac{A_\cC}{\ell_c^2}.}}

We note that a  calculation of the Rindler and Schwarzschild horizons with similar boundary conditions should in
principle be  { possible}
if
one employs a suitable  radial IR cut-off  to regulate the radial $\cL^2$ norm,  so that Eqn.~\eqref{redssee.eq} is well defined.

\section{Discussion}
\label{discussion.sec} 

In this work we  began with an analysis of  the SSEE, using the two sets of modes in $\cM$ and $\cO
\subset \cM$.  We found that  when the Bogoliubov
transformations satisfy certain conditions in both the finite as well as the static, spatially finite cases, there
are real solutions to the eigenvalue equations which come in pairs $(\mu, 1-\mu)$. We then calculated the SSEE
for de Sitter horizons in $d>2$ as well as Schwarzschild de Sitter horizons in $d>2$.
We found that in both cases, the eigenvalues also satisfy the condition  $\mu\not\in (0,1)$, as
expected from the arguments given in \cite{ssee}.  In both spacetimes, we used the cut-off in the angular
modes to demonstrate that  $\cS \propto A$ for $d>2$.  This is as expected, and is a further confirmation that the SSEE
is a good measure of  entanglement entropy. 

When we restrict to  $d=2$, however, we find that the SSEE is constant and thus  {\it not}  of  the Calabrese-Cardy form. This
differs from the results of earlier $d=2$ calculations of the SSEE  both in the continuum and using causal set discretisations
\cite{saravani2014spacetime,Mathur:2021zzl,Sorkin:2016pbz,Belenchia:2017cex,Surya_2021}, where the Calabrese-Cardy form
was obtained. We note that this is not a
feature only of the SSEE alone but also of the associated Von Neumann EE in $d=2$, and follows from an extension of the results of \cite{Higuchi:2018tuk} to $d=2$. 

An obvious difference with earlier calculations is that $\cO$ in the de Sitter cases studied here are non-compact. For
the  nested causal diamonds in $d=2$ Minkowski spacetime as well as the causal diamond on the finite cylinder spacetime, $\cO$ is chosen to be the domain of dependence of a finite interval, and is therefore compact \cite{saravani2014spacetime,Mathur:2021zzl}.  In de Sitter spacetime, the domain of dependence of the half circle is the static patch which is not compact. We have shown  that despite the temporal non-compactness, the SSEE
equation Eqn.~\eqref{ssee.eq} is well defined for the static patch. On the other hand, the numerical calculation for de Sitter causal sets
\cite{Surya_2021} necessitated an  IR cut-off, so that the regions $(\cM,g)$ as well as $\cO$ differ from those used in
this work. After a suitable truncation in the discrete spectrum,  the  Calabrese-Cardy form for the
causal set SSEE was recovered.  Technically, one of the features that simplified our calculations was the diagonal form
Eqn.~\eqref{diag.eq},\eqref{2ddiag.eq}, which, as we had noted in Sec.~\ref{gf.sec}, is not satisfied for the $d=2$
cylinder calculation of \cite{Mathur:2021zzl}. Re-examining our calculation we see that a temporal IR cut-off in  $\cO$
would destroy this diagonal property. Whether this could restore the logarithmic behaviour or not would be difficult to
establish analytically, but given the causal set example, it suggests that this may indeed be the case.  This in turn
suggests new subtleties in the nature of $d=2$ entanglement in curved spacetime, which should be explored.  

We also note that in these calculations, the angular modes tranform trivially. Thus, the 
generalised eigenvalues are dimension independent, which makes the $d=2$ calculation a simple dimensional
restriction. Hence the conclusions we draw in higher dimensions -- namely that $\cS$ has an area dependence -- also implies that the SSEE is constant in $d=2$.  In higher dimensions the density
of states comes from the degeneracy of the angular modes on $\ms^{d-2}$  which necessitates a cut-off, while that 
in $d=2$ comes from the two ``angular modes'' on $\ms^0$.

Ultimately, the use of the SSEE lies in its covariant formulation and its applicability to systems where Hamiltonian
methods are not at hand. This is the case with causal set quantum gravity, since the analogues of spatial hypersurfaces
allow for  a certain ``leakage''  of information.  As shown in \cite{Sorkin:2016pbz,Surya_2021} the calculation of the SSEE for QFT on
causal sets throws up some unexpected behaviour, due to the non-local but covariant nature of the UV cut-off.  It is of
course not obvious that EE plays a fundamental role in
quantum gravity, but the effects of the latter can be non-trivial when discussing emergent phenomena.

The SSEE approach to EE  is compatible with that of  algebraic quantum field theory, where entanglement measures are
state functionals which measure  the  entanglement of a  mixed state $\hW_{\cO}$ obtained by restricting the pure
state $\hW$ in $\cM \supset \cO$. The  SSEE was motivated by the study of systems with finite degrees of freedom, but
has been shown to give the expected results for systems with infinite degrees of freedom, as is the case here and the
$d=2$ examples discussed above. Defining EE for systems with infinite degrees of
freedom is however known to be non-trivial;  type III  algebras which characterise QFT  do  not
factor, thus  leading  to significant  complications  (see \cite{Hollands:2017dov}).   Although we
have several QFT examples for  which the SSEE is a good entanglement measure, an important open  question is whether it
can be rigorously derived using methods from algebraic quantum field
theory.  

\section*{Acknowledgements}
SS is supported in part by a Visiting Fellowship at the Perimeter Institute. Research at Perimeter Institute is supported in part by the Government of Canada through the  Department  of  Innovation,  Science  and  Economic  Development  Canada  and  by  the Province of Ontario through the Ministry of Colleges and Universities.

\appendix

\section{SSEE of the $\alpha$ vacua for the dS static patch}\label{alphabeta.sec}

In this section, we compute the SSEE for the $\alpha$ vacua \cite{Mottola:1984ar,Allen:1985ux} $\bv_{klm}^{(\alpha,\beta)}\equiv\varphi^{(\alpha,\beta)}_{kl}(\eta,r)Y_{lm}(\theta,\phi)$ which can be parameterized as 
\be
\varphi^{(\alpha,\beta)}_{kl}(\eta,r)\equiv \cosh(\alpha)\varphi_{kl}(\eta,r)+\sinh(\alpha)e^{i\beta}\varphi^*_{kl}(\eta,r),
\label{alphabetabdmodes}
\ee
where $\alpha\in[0,\infty)$ and $\beta\in(-\pi,\pi)$. 
Expressing these in  terms of the $\bu_{plm}$ modes in $\Righf$ as
\eq{
\varphi^{(\alpha,\beta)}_{kl}(\eta,r) = \int_0^\infty dp\, \left(\alpha^{(\alpha,\beta)}_{kp}\psi_{pl}(t,x) + \beta^{(\alpha,\beta)}_{kp}\psi_{pl}^*(t,x)\right),
}
where
\eq{
\alpha^{(\alpha,\beta)}_{kp}=\frac{1}{n_p}\left<\bu_{plm},\bv^{(\alpha,\beta)}_{klm}\right>_\Righf\quad\text{and}\quad \beta^{(\alpha,\beta)}_{kp}=\frac{1}{n_p}\left<\bu_{pl-m}^*,\bv^{(\alpha,\beta)}_{klm}\right>_\Righf.
}
Using Eqn.~\eqref{alphabetabdmodes}, \eqref{eq:pqip}, \eqref{eq:alpkp} and \eqref{eq:betkp}, we find the coefficients $\alpha^{(\alpha,\beta)}_{kp}$ and $\beta^{(\alpha,\beta)}_{kp}$ to be
\eq{
\alpha^{(\alpha,\beta)}_{kp} = \cosh(\alpha)\alpha_{kp} + \sinh(\alpha)e^{i\beta}\beta_{kp}^* = \frac{2^{-ip}k^{ip-\frac{1}{2}}}{\sqrt{2\pi(1-e^{-2\pi p})}}\left(\cosh(\alpha)+e^{-\pi p}e^{i\beta}\sinh(\alpha)\right),\\
\beta^{(\alpha,\beta)}_{kp} = \cosh(\alpha)\beta_{kp} + \sinh(\alpha)e^{i\beta}\alpha_{kp}^* = \frac{2^{ip}k^{-ip-\frac{1}{2}}}{\sqrt{2\pi(1-e^{-2\pi p})}}\left(e^{-\pi p}\cosh(\alpha)+e^{i\beta}\sinh(\alpha)\right).
}
Further calculation shows that $A^{(\alpha,\beta)}_{pp'} \equiv \int dk\, \alpha^{(\alpha,\beta)}_{kp}\alpha^{(\alpha,\beta)*}_{kp'}$, $B^{(\alpha,\beta)}_{pp'} \equiv \int dk\, \alpha^{(\alpha,\beta)}_{kp}\beta^{(\alpha,\beta)*}_{kp'}$, $C^{(\alpha,\beta)}_{pp'} \equiv \int dk\, \beta^{(\alpha,\beta)}_{kp}\alpha^{(\alpha,\beta)*}_{kp'}$ and $D^{(\alpha,\beta)}_{pp'} \equiv \int dk\, \beta^{(\alpha,\beta)}_{kp}\beta^{(\alpha,\beta)*}_{kp'}$ is of the form
\eq{
A^{(\alpha,\beta)}_{pp'} = a^{(\alpha,\beta)}_p\delta(p-p'),\;D^{(\alpha,\beta)}_{pp'} = d^{(\alpha,\beta)}_p\delta(p-p'),\;B^{(\alpha,\beta)}_{pp'} =C^{(\alpha,\beta)}_{pp'} =0,
}
where
\eq{
	a^{(\alpha,\beta)}_p &=\frac{1}{1-e^{-2\pi p}}\left(\cosh^2(\alpha)+e^{-2\pi p}\sinh^2(\alpha)+e^{-\pi p}\sinh(2\alpha)\cos(\beta)\right),\\
	d^{(\alpha,\beta)}_p &=\frac{1}{1-e^{-2\pi p}}\left(e^{-2\pi p}\cosh^2(\alpha)+\sinh^2(\alpha)+e^{-\pi p}\sinh(2\alpha)\cos(\beta)\right).
}

\begin{figure}[h!]
\centering\includegraphics[height=5cm]{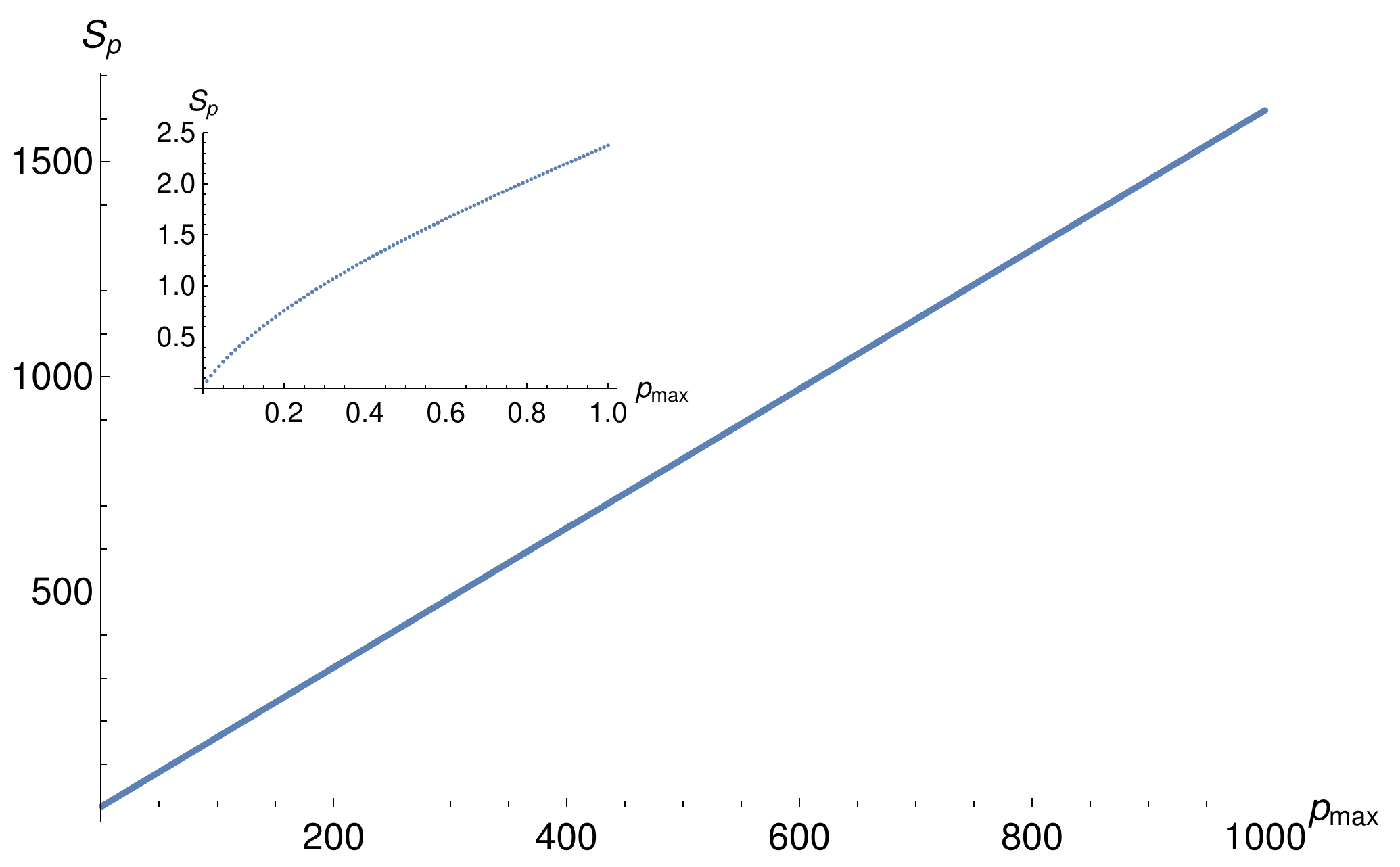}
\caption{A plot of $S_p$ vs $p_\mx$ for $(\alpha,\beta)=(1,0)$. We see that for large enough $p_\mx$, $S_p\propto p_\mx$.}
\label{fig:sponezero}
\end{figure}

Generalised eigenvalues $\mu$ is then
\eq{
    \mu^{+(\alpha,\beta)}_{p} = a^{(\alpha,\beta)}_p\quad\text{and}\quad \mu^{-(\alpha,\beta)}_{p} =-d^{(\alpha,\beta)}_p,
}
from which we obtain the SSEE as
\eq{
S^{(\alpha,\beta)} &= (l_\mx+1)^2\int_0^\infty dp\left(a^{(\alpha,\beta)}_p\log(a^{(\alpha,\beta)}_p) - d^{(\alpha,\beta)}_p\log(d^{(\alpha,\beta)}_p)\right).\label{eq:sseeinstatic}
}
Unlike for the SSEE obtained from the Bunch-Davies vacuum $(\alpha=0)$, $S^{(\alpha,\beta)}$ is in general dependent on the cut-off in $p$. As an example, for $(\alpha,\beta)=(1,0)$, we evaluate the integral in Eqn.~\eqref{eq:sseeinstatic} numerically for different cut-offs in $p$ and find that the SSEE depends on both $p_\mx$ and $l_\mx$, and is of the form
\be
S^{(1,0)}= S_p(p_\mx)(l_\mx+1)^2,
\ee
where for large enough $p_\mx$, $S_p(p_\mx)$ is found to be proportional to $p_\mx$ as shown in Fig.~\ref{fig:sponezero}.
\section{SSEE of general $d$-dimensional dS horizon}\label{ddimds.sec}
In this section, we extend our calculation of SSEE in four dimensional de Sitter to a general $d$-dimensional de Sitter spacetime with $d>2$ and show that the entropy depends on the spacetime dimension solely due to the dimension dependent degeneracy of the spherical harmonics.

We start with showing that the Bunch-Davies modes $\{\bv_{kL}\}$ in the conformal patch of $d$-dimensional de Sitter spacetime is given by $\bv_{kL}=\varphi_{kl}(\eta,r)Y_L(\Omega_{d-2})$, where
\begin{equation}
\varphi_{kl}(\eta,r) = \frac{H^{d/2-1}}{\sqrt{2k}}(-k\eta)^{\frac{d-1}{2}}H^{(1)}_\nd(-k\eta)(kr)^{2-d/2}j_{l+\frac{d}{2}-2}(kr),\;k>0.\label{eq:ddimbdmodes}
\end{equation}
Here $L$ represents a collection of indices $\{l,l_1,\ldots,l_{d-4},m\}$ such that $l,l_1,\ldots l_{d-4} \in \{0,1,2,\ldots\}$, $m\in\mathbb{Z}$ and $l\geq l_1\geq \ldots l_{d-4}\geq |m|$. $\Omega_{d-2}$ represents a collection of angular coordinates on $\mathbb{S}^{d-2}$. We can clearly see that for $d=4$, these modes reduces to the Bunch-Davies modes given by Eqn.~\eqref{eq:bdmodes}. For $\{\bv_{kL}\}$ to qualify for the QFT modes they have to be Klein-Gordon orthonormal solutions of the Klein-Gordon equation, which we will show now.

Klein-Gordon equation for the massive scalar field with effective mass $\mass$ in de Sitter spacetime is
\begin{equation}
-\eta^d\partial_\eta(\eta^{2-d}\partial_\eta\phi) + \frac{\eta^2}{r^{d-2}}\partial_r(r^{d-2}\partial_r\phi) + \frac{\eta^2}{r^2}\nabla_{\Omega_{d-2}}^2\phi = \frac{\mass^2}{H^2}\phi. \label{eq:ddimkgeqn}
\end{equation}
For $\varphi_{kl}$ given by Eqn.~\eqref{eq:ddimbdmodes} we find
\begin{eqnarray}
-\eta^d\partial_\eta(\eta^{2-d}\partial_\eta\varphi_{kl}) &=& \frac{1}{4}\left(1+d(d-2)+4k^2\eta^2-4\nu^2\right)\varphi_{kl},\\
\frac{\eta^2}{r^{d-2}}\partial_r(r^{d-2}\partial_r\varphi_{kl}) &=& \frac{\eta^2}{r^2}\left(l(l+d-3)-k^2r^2\right)\varphi_{kl},\\
\frac{\eta^2}{r^2}\nabla_{\Omega_{d-2}}^2 Y_L &=& -\frac{\eta^2}{r^2}l(l+d-3)Y_L.
\end{eqnarray}
Therefore $\varphi_{kl}$ given by Eqn.~\eqref{eq:ddimbdmodes} solves the Klein-Gordon equation for
\begin{equation}
\nd = \sqrt{\left(\frac{d-1}{2}\right)^2-\frac{\mass^2}{H^2}}.
\end{equation}
The modes $\{\bv_{kL}\}$ are Klein-Gordon orthonormal: 
\begin{equation}
\begin{split}
(\bv_{kL},\bv_{k'L'})_{\Righf\cup\fut} = i\frac{kk'}{2}|\eta|\left(H_\nd^{{(1)}*}(-k\eta)\partial_\eta H_\nd^{{(1)}}(-k'\eta) - H_\nd^{{(1)}}(-k'\eta)\partial_\eta H_\nd^{{(1)}*}(-k\eta)\right)\\
\int_0^\infty dr\,r^2 j_{l+\frac{d}{2}-2}(kr)j_{l'+\frac{d}{2}-2}(k'r)\int_{\mathbb{S}^{d-2}} d\Omega_{d-2}Y_L^*(\Omega_{d-2})Y_{L'}(\Omega_{d-2}).
\end{split}
\end{equation}
Here the volume element on the constant $\eta$ surface $\spac$ is $d\spac = (H\eta)^{1-d}r^{d-2}dr d\Omega_{d-2}$ and the future pointing unit vector normal to $\spac$ is $\hn^\mu\partial_\mu = H|\eta|\partial_\eta$. Using the fact that spherical harmonics are $L^2$ orthonormal on $\mathbb{S}^{d-2}$ and
\begin{equation}
\int_0^\infty dr\,r^2 j_{n}(kr)j_{n}(k'r) = \frac{\pi}{2k^2}\delta(k-k'),
\end{equation}
for $n>-1$, we can write
\begin{equation}
(\bv_{kL},\bv_{k'L'})_{\Righf\cup\fut} = \frac{i\pi}{4}|\eta| \left(H_\nd^{{(1)}*}(-k\eta)\partial_\eta H_\nd^{{(1)}}(-k\eta) - H_\nd^{{(1)}}(-k\eta)\partial_\eta H_\nd^{{(1)}*}(-k\eta)\right)\delta(k-k')\delta_{LL'}.\label{eq:kgbdf}
\end{equation}
Since the Klein-Gordon inner product is independent of the choice of the spacelike hypersurface, we will evaluate it at the surface $\eta\rightarrow-\infty$, where 
\begin{equation}
H^{(1)}_\nd(-k\eta) \rightarrow \sqrt{\frac{-2}{\pi k\eta}}e^{-i\left(k\eta+\frac{\pi\nd}{2}+\frac{\pi}{4}\right)}.\label{eq:hankelapprox}
\end{equation}
Substituting Eqn.~\eqref{eq:hankelapprox} in Eqn.~\eqref{eq:kgbdf}, we see that
\begin{equation}
(\bv_{kL},\bv_{k'L'})_{\Righf\cup\fut} = \delta(k-k')\delta_{LL'}.
\end{equation}
Similarly we can show that
\begin{equation}
(\bv_{kL}^*,\bv_{k'L'}^*)_{\Righf\cup\fut} = -\delta(k-k')\delta_{LL'}\quad\text{and}\quad (\bv_{kL},\bv_{kL}^*)_{\Righf\cup\fut} = 0.
\end{equation}

As in the case of $d=4$, we show that in region $\Righf$, we have a Klein-Gordon orthonormal set of modes given by $\bu_{pL}=\psi_{pl}(t,x)Y_L(\Omega_{d-2})$, where
\eq{
\psi_{pl}(t,x) =\sqrt{2\sinh(\pi p)}\,\npl^{(d,\nd)} \,U_{pl}^{(d,\nd)}(x)e^{-ipt},\quad p>0, \label{eq:stat_modes_ddim}
}
with
\eq{
U_{pl}^{(d,\nd)}&=x^l(1-x^2)^{\frac{-ip}{2}}{_2F_1}\left(\frac{\frac{d-1}{2}+l-ip+\nd}{2},\frac{\frac{d-1}{2}+l-ip-\nd}{2},l+\frac{d-1}{2};x^2\right)\nonumber\\
&= x^{2-\frac{d}{2}}U_{p\tl}^{(4,\nd)}(x),
\label{eq:Upl}\\
\npl^{(d,\nd)} &= \frac{H^{\frac{d}{2}-1}}{2\sqrt{2}\pi\Gamma\left(l+\frac{d-1}{2}\right)}\Gamma\left(\frac{l+\frac{d-1}{2}-ip+\nd}{2}\right)\Gamma\left(\frac{l+\frac{d-1}{2}-ip-\nd}{2}\right)\nonumber\\
&=H^{\frac{d}{2}-2}N_{p\tl}^{(4,\nd)},\label{eq:Npl}
}
where $\tl = l+\frac{d}{2}-2$, and the $U_{p\tl}^{(4,\nd)}(x)$ and $ N_{p\tl}^{(4,\nd)}$ carry the extra label
$\nu_d(\mass) \neq \nu_4(\mass)$. The  Klein-Gordon inner product  
\begin{eqnarray}
(\bu_{pL},\bu_{p'L'})_{\Righf} &=& 2(p+p')H^{2-d}\sqrt{\sinh(\pi p)\sinh(\pi p')}e^{i(p-p')t}\npl^{(d,\nd)*}N_{p'l}^{(d,\nd)}\nonumber\\
&&\times\int_0^1 dx\,\frac{x^{d-2}}{1-x^2}U_{pl}^{(d,\nd)}(x)U_{p'l}^{(d,\nd)}(x)\delta_{LL'},\label{eq:kgstatddim}
\end{eqnarray}
where $d\spac=H^{1-d}(1-x^2)^{-1/2}x^{d-2}$  on  the Cauchy hypersurface $\spac_t$ and the future
pointing unit vector normal to the $\spac$ is $\hn^\mu\partial_\mu = H(1-x^2)^{-1/2}\partial_t$. Using the relations Eqn.~\eqref{eq:Upl} and \eqref{eq:Npl}, we see that the $\{\bu_{pL}\}$ are Klein-Gordon orthogonal as in the $d=4$ case,
\begin{equation}
(\bu_{pL},\bu_{p'L'})_{\Righf} = \delta(p-p')\delta_{LL'}.
\end{equation}
We can similarly show that
\begin{equation}
(\bu_{pL}^*,\bu_{p'L'}^*)_{\Righf} = -\delta(p-p')\delta_{LL'}\quad\text{and}\quad (\bu_{pL}^*,\bu_{p'L'})_{KG} = 0.
\end{equation}
Using 
\eq{
\int_0^\infty dk\; k^{-ip-\frac{1}{2}}\varphi_{kl}(\eta,r) = 2^{-ip}e^\frac{\pi p}{2}\npl^{(d,\nd)}U_{pl}^{(d,\nd)}(x)e^{-ipt},\label{eq:phetouint}
} 
we see that the Bogoliubov transformation between $\{\bv_{kL}\}$  and $\{\bu_{pL}\}$  in  $\Righf$ are given by
Eqn.~\eqref{eq:alpkp} and  ~\eqref{eq:betkp} and are the same for all dimensions. This immediately implies that the
mode-wise entropy is given by Eqn.~\eqref{eq:hy}, with an infinite degeneracy coming from the angular modes. Integrating
over $p \in (0, \infty)$ gives us a finite answer as before, but we need to impose an angular cut-off  $l_\mx$ as we did in
$d=4$. The regulated  SSEE  is then 
\eq{ \cS &= \sum_{L, l=0}^{l_\mx} \frac{\pi}{6} =
  \frac{\pi}{6}\frac{(2l_\mx+d-2)(l_\mx+d-3)!}{l_\mx!(d-2)!} \nonumber \\
  & \simeq \frac{\pi}{6} \frac{2}{(d-2)!} l_\mx^{d-2}, \quad l_\mx>>1.}
As in $d=4$ 
using the approximate flatness of the metric at the equator, $d\Omega \simeq (2\pi/l_\mx)^{d-2} \simeq
(l_c H)^{d-2}$,  which means that $ \cS \propto \frac{A_c}{l_c^{d-2}}$.  

\bibliography{reference}
\bibliographystyle{ieeetr}
\end{document}